\documentclass[conference]{IEEEtran}
\RequirePackage{etex}
\usepackage{cite}
\usepackage{amsmath,amssymb,amsfonts,amsthm}
\usepackage{algorithm}
\usepackage{romannum}
\usepackage{graphicx}
\usepackage{textcomp}
\usepackage{xcolor}
\usepackage{algpseudocode}
\usepackage{listings}
\usepackage{booktabs}
\usepackage{tabularx}
\newcolumntype{L}{>{\raggedright\arraybackslash}X}
\usepackage{textcomp}
\usepackage{caption}
\usepackage{subcaption}
\usepackage{svg}
\usepackage{adjustbox}
\usepackage{xspace}
\usepackage{multirow}
\usepackage{etoolbox}
\usepackage{svg}
\usepackage{tcolorbox}
\usepackage{linegoal}
\usepackage{threeparttable}
\usepackage{comment}
\usepackage{booktabs}

\usepackage{cleveref}
\usepackage{float}
\newtcolorbox{mybox}{colback=blue!10!white,colframe=green!0!white}
\newtheoremstyle{mystyle}
  {}
  {}
  {}
  {}
  {\bfseries}
  {:}
  { }
  {\thmname{#1}\thmnumber{ #2}\thmnote{ (#3)}}
\theoremstyle{mystyle}
\newtheorem{finding}{Finding}

\graphicspath{ {./images/} }
\usepackage{soul}
\def\BibTeX{{\rm B\kern-.05em{\sc i\kern-.025em b}\kern-.08em
    T\kern-.1667em\lower.7ex\hbox{E}\kern-.125emX}}
\makeatletter
\newcommand{\linebreakand}{%
  \end{@IEEEauthorhalign}
  \hfill\mbox{}\par
  \mbox{}\hfill\begin{@IEEEauthorhalign}
}
\makeatother

\definecolor{amber}{rgb}{1.0, 0.75, 0.0}

\begin{document}

\title{YFlows: Systematic Dataflow Exploration and Code Generation for Efficient Neural Network Inference using SIMD Architectures on CPUs\\}
\author{\IEEEauthorblockN{Cyrus Zhou}
\IEEEauthorblockA{\textit{Department of Computer Science} \\
\textit{University of Chicago}\\
Chicago, IL \\
zhouzk@uchicago.edu}
\and
\IEEEauthorblockN{Zachary Hassman}
\IEEEauthorblockA{\textit{Department of Computer Science} \\
\textit{University of Chicago}\\
Chicago, IL \\
zhassman@uchicago.edu}
\and
\IEEEauthorblockN{Ruize Xu}
\IEEEauthorblockA{\textit{Department of Computer Science} \\
\textit{University of Chicago}\\
Chicago, IL \\
richard1xur@uchicago.edu}
\linebreakand 
\IEEEauthorblockN{Dhirpal Shah}
\IEEEauthorblockA{\textit{Department of Computer Science} \\
\textit{University of Chicago}\\
Chicago, IL \\
dhirpalshah@uchicago.edu}
\and
\IEEEauthorblockN{Vaughn Richard}
\IEEEauthorblockA{\textit{Department of Computer Science} \\
\textit{University of Chicago}\\
Chicago, IL \\
vaughnrichard@uchicago.edu}
\and
\IEEEauthorblockN{Yanjing Li}
\IEEEauthorblockA{\textit{Department of Computer Science} \\
\textit{University of Chicago}\\
Chicago, IL \\
yanjingl@uchicago.edu}
}
\maketitle

\begin{abstract}
We address the challenges associated with deploying neural networks on CPUs, with a particular focus on minimizing inference time while maintaining accuracy. Our novel approach is to use the dataflow (i.e., computation order) of a neural network to explore data reuse opportunities using heuristic-guided analysis and a code generation framework, which enables exploration of various Single Instruction, Multiple Data (SIMD) implementations to achieve optimized neural network execution. Our results demonstrate that the dataflow that keeps outputs in SIMD registers while also maximizing both input and weight reuse consistently yields the best performance for a wide variety of inference workloads, achieving up to 3x speedup for 8-bit neural networks, and up to 4.8x speedup for binary neural networks, respectively, over the optimized implementations of neural networks today.
\end{abstract}


\section{Introduction}\label{sec:intro}
In recent years, neural networks have expanded their reach beyond high-performance computing environments, permeating low-end servers and edge devices such as smartphones, IoT devices, and smart sensors \cite{bib:edge,hadidi2019characterizing,shi2016edge,lane2016deepx}. However, the deployment of neural networks on these devices presents various challenges, with inference time being a critical factor \cite{lane2016deepx, sze2017efficient, howard2017mobilenets, iandola2016squeezenet, han2016eie}. The Single Instruction, Multiple Data (SIMD) capabilities of contemporary CPUs present an opportunity to accelerate neural networks. SIMD allows a single instruction to be executed on multiple data elements concurrently, thereby substantially improving computational throughput and overall performance, and yielding benefits in terms of both energy conservation and efficient utilization of computational resources \cite{hennessy2011computer, lee2018efficient,pu2011xetal}. 

\textit{Dataflow} refers to an execution order of computational operations of a neural network, and it is an important consideration when utilizing SIMD for inference. It determines the reuse opportunities of different variables (e.g., inputs, weights, and outputs), and can therefore guide how to best allocate valuable SIMD register resources to maximize reuse. While dataflows for deep learning accelerators have been extensively explored \cite{bib:szedataflow, bib:pardataflow, bib:diannao, bib:handataflow}, the majority of previous studies and libraries for CPUs do not consider dataflows \cite{tensorflow2015,pytorch2019,bib:tvm,bib:bitflow}. Instead, weight stationary, i.e., keep using the same weight value until all computations requiring this value are done before moving on to the next weight value, is widely adopted \cite{bib:liuatc, bib:onednn, larq}. However, we found that by adopting the carefully designed dataflow and co-optimizing with other techniques (i.e., blocking, operator fusion), the inference speed can be improved significantly, up to 3.5 times, compared to state-of-the-art implementations of 8-bit integer networks \cite{bib:tvm}, and $>$10 / 4.8 times compared to optimized bitserial \cite{bib:tvm,cowan20automatic} / state-of-the-art SIMD \cite{bib:liuatc} implementations of binary neural networks, respectively.

Unfortunately, compiler support for efficient SIMD code generation is lacking \cite{amiri2020simd,govindaraju2013breaking,larsen2000exploiting}, as demonstrated in our experiments on x86 and ARM architectures. Programs written to explicitly utilize SIMD often receive no further compiler optimization, such as harnessing unused vector registers \cite{liu2023minotaur}. Furthermore, auto-vectorization features in compilers \cite{arm2023neon,intel2023intrinsics} overlook vectorizable scalar implementations \cite{amiri2020simd,mendis2018goslp,govindaraju2013breaking,kim2012efficient}, possibly due to the expansive search space noted in \cite{xiao2022structural}.

The nuances of SIMD optimization, such as ensuring non-dependency in vector register values, are highlighted in \cite{kim2012efficient,liu2023minotaur}. These complexities are compounded by the reliance on fragile heuristics in current autovectorization techniques, as critiqued in \cite{mendis2018goslp,haj2020neurovectorizer,larsen2000exploiting}. This is also true for highly optimized frameworks like TVM \cite{bib:tvm} as they rely on compiler backends such as LLVM \cite{bib:llvm}. With these challenges, the burden of SIMD optimization predominantly lies with programmers. Consequently, there's a pressing need for a systematic approach to maximize SIMD implementation efficiency. 
 
To this end, we present the first work that employs the notion of dataflow to systematically explore the full SIMD computation capacities on CPUs for efficient neural network inference. The major contributions include:
\begin{enumerate}
    \item{We extended the existing dataflows, which typically specify only one type of variable to be reused, by allowing all types of variables to be reused. Extended dataflows enable systematic exploration to fully utilize SIMD register resources, and substantially reduce costs associated with data and instruction movements.}
    \item{We formalized a set of heuristics, based on data movement costs, to optimize three basic, general neural network dataflows - defined in Sec. \ref{sec:bg-dataflows} - by maximizing reuse opportunities within each dataflow.
    }
    \item{We implemented a code generator that automatically uses SIMD instructions to implement the three basic dataflows and various extended dataflows, for any given neural network configuration. This code generator allows us to compare different dataflows to determine the most efficient implementation.}
    \item{We quantitatively compared our best implementation against state-of-the-art implementations using representative workloads, and show that our results achieve substantial improvements: up to 3.5x speedup for 8-bit neural networks (against TVM \cite{bib:tvm}), and up to 4.8x speedup for binary neural networks (against \cite{bib:liuatc}), respectively.}
\end{enumerate}


\section{Basic Dataflows of Neural Networks} \label{sec:bg-dataflows}

Three major, basic dataflows have been identified in the literature\footnote{We exclude dataflows that are specifically tailored to specific deep learning accelerator architectures (e.g., \textit{Row-starionary} \cite{bib:szedataflow}, \textit{No-local-reuse} \cite{bib:fusedacc}, etc.) as they cannot be applied to CPUs. For example, row-stationary keeps software variables stationary in the rows of processing engines of a 2D systolic array; however, there is no notion of ``rows of cores" in CPUs.}\cite{kwon2019understanding,yang2020procrustes,samajdar2018scale}, as shown in  Algorithms \ref{df-algo-is}, \ref{df-algo-ws}, and \ref{df-algo-os} in the semantics of ARM SIMD intrinsics \cite{ArmIntrinsics}, using convolution layers as an example.

\subsection{Input Stationary (IS)}\label{sec:bg-dataflows-is}
IS operates by iterating through the input tensor. It applies all relevant filters to each input and accumulates the results to the respective entries in the output.

\begin{algorithm}
\caption{IS Dataflow for Convolution Layers.}
\label{df-algo-is}
\small
\begin{algorithmic}
\Require inputs[$H$], weights[$R$], outputs[$E$]
\For{$h$ in $H$}
    \State input $\gets$ $vload$(\&inputs[$h$]);
    \For{$r$ in $R$}
        \State weight = $vload$(\&weights[$r$]);
        \State calculate $e$ from $h$, $r$;
        \State outputs[$e$] += $vredsum$($vmul$(input, weight));
    \EndFor
\EndFor
\end{algorithmic}
\end{algorithm}

\subsection{Weight Stationary (WS)}\label{sec:bg-dataflows-ws}
WS iterates through the weight tensor. For each output entry whose computation depends on the current weight tensor, WS collects each relevant entry from the input for computations and accumulates the result to the corresponding output. 

\begin{algorithm}
\caption{WS Dataflow for Convolution Layers.}
\small
\begin{algorithmic}
\Require inputs[$H$], weights[$R$], outputs[$E$]
\For {$r$ in $R$}
    \State weight $\gets$ $vload$(\&weights[$r$]);
    \For {$e$ in $E$}
        \State calculate $i$ from $e$, $r$;
        \State input = $vload$(\&inputs[$i$]);
        \State outputs[$e$] += $vredsum$($vmul$(input, weight));
    \EndFor
\EndFor
\end{algorithmic}
\label{df-algo-ws}
\end{algorithm}

\subsection{Output Stationary (OS)}\label{sec:bg-dataflows-os}
OS iterates through the output tensor. It performs all necessary multiply-accumulate computations to obtain the final result for one output entry before moving on to the next.

\begin{algorithm}
\caption{OS Dataflow for Convolution Layers.}
\small
\begin{algorithmic}
\Require inputs[$H$], weights[$R$], outputs[$E$]
\For {$e$ in $E$}
    \State output = $vmov$($\vec{0}$)
    \For {$r$ in $R$}
        \State if such $i$ exists, calculate $i$ from $e$, $r$, else continue;
        \State input, weight  = $vload$(\&inputs[$i$]), $vload$(\&weights[$r$]);
        \State output = $vadd$($vmul$(input, weight),output);
    \EndFor
    \State outputs[$e$] = $vredsum$(output);
\EndFor
\end{algorithmic}
\label{df-algo-os}
\end{algorithm}

\subsection{Memory layout and Computation Order}
\begin{figure}
\includegraphics[width=\linewidth]{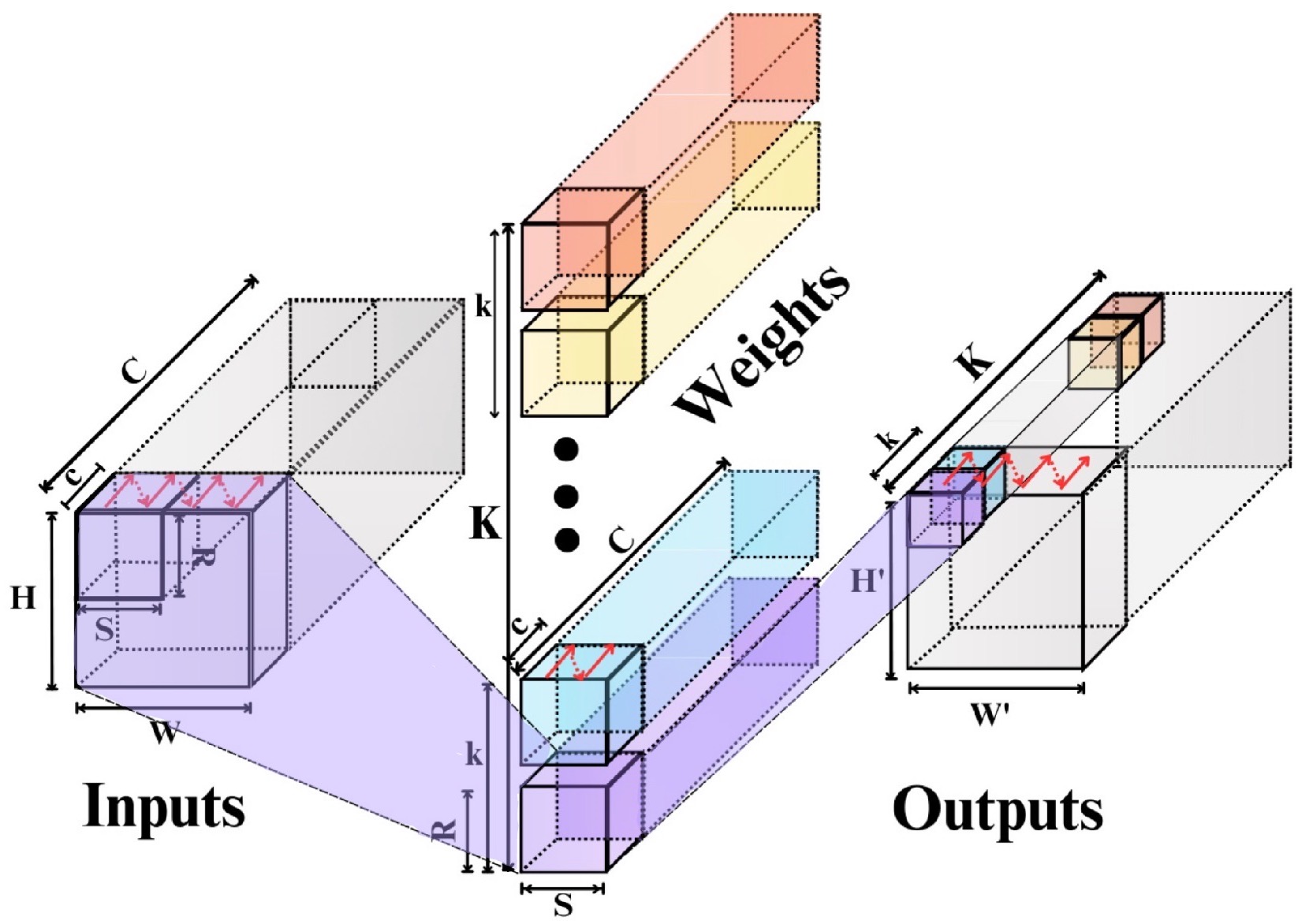}
\caption{Memory layout of tensors. Red arrows show a subset of data elements following sequential memory addresses. Input channel blocks are traversed first along the output channel dimension. The purple shade covers a single vector variable. }
\label{fig:mem-alignment}
\end{figure}

Naturally, the computation order under a dataflow follows the sequential memory addresses of the corresponding data elements. We illustrate the memory layout scheme in Fig. \ref{fig:mem-alignment}.

We opt for the NCHW[xc] memory layout for each input/output tensor. In traditional NCHW alignment, tensors are arranged by first the number of images (batch size, N), then channels (C), followed by height (H), and lastly width (W). In NCHW[xc], data are grouped into blocks of size $x \times H \times W$, and we call these blocks \emph{channel blocks}. 
The channel blocks follow the NCHW layout, while data in each channel block follows the HW[xc] layout, and $x$ is typically chosen so that $x \times element\_width$ is a multiple of the size of the physical vector registers ($1\text{-}3\times$ in our implementation).

There are two main reasons for this memory layout choice. First, vectorization in the channel dimension streamlines vector computations, avoiding excessive operations such as shifting, because the number of channels multiplied by data size in a neural network layer is usually a multiple of SIMD register length (or vice versa).  Previous works have demonstrated the effectiveness of this scheme for floating-point, integer and binary neural networks \cite{bib:liuatc, bib:directconv, bib:bitflow}.

Second, NCHW[xc] enables data reuse between successive channel blocks. With NHWC, no element engages in calculations for two successive elements, whether inputs, weights, or outputs, under any dataflow. In contrast, NCHW[xc] enables various dataflows to be exploited to maximize data reuse (see Sec. \ref{sec:ext-df}). Note that, for binary networks, NHWC can be largely the same as NCHW[xc] in performance since the number of channels in most network architectures is $\leq 512$ and a multiple of vector register size in modern ISAs \cite{bib:bitflow}.

To optimize weight data access locality, we adopt the CKRS[xc] memory layout (matching the input/output tensor layout), where $C$, $K$, $R$, $S$ denote $\#$Input Channels, $\#$Output Channels, $\#$rows/filter height, $\#$columns/filter width, respectively, and $x$ in the notation for the weight tensor is chosen to be exactly the $x$ of the input tensor. Following this layout, the output tensors can be written back sequentially regardless of the size of the input/output channel blocks and dataflows. 

In terms of the compute order across input channel blocks, for better memory locality (as validated by our observation), we proceed along the output channel dimension before moving onto the next input channel block. In other words, the loop on the input channel dimension is an outer loop of that on the output channel dimension. 


\subsection{Implementation and Performance of Basic SIMD Dataflows}
In software, we declare three \textit{vector variables} to implement any of the three basic dataflows, one for each of the input, weight, and output data types. 
The size of each vector variable is $x \times element\_width$ as shown in Fig.~\ref{fig:mem-alignment}), which is a multiple of the vector register size. Also, the total size of all vector variables is less than or equal to the total size of all vector registers. We distinguish these two terms because physical vector registers in some architectures can be concatenated to form longer vectors. For example, in ARM, vector registers are $128$ bits in size, but vector variables can be multiples of $128$ bits occupying multiple physical registers.

\begin{figure}
\includegraphics[width=\linewidth]{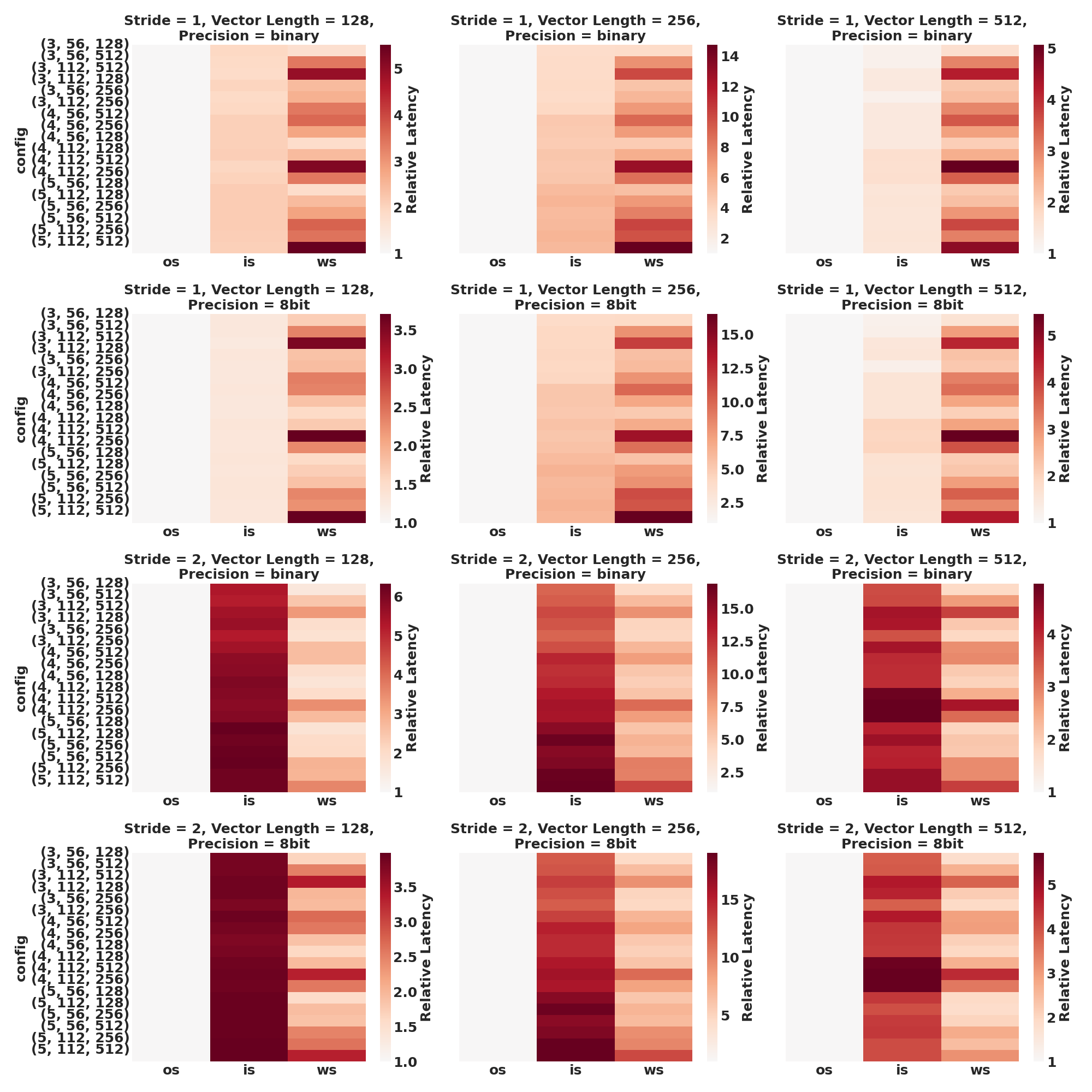}
\caption{Relative latency of basic dataflows for various convolution layers for $\text{Vector Length} = (elem\_width \times c) \in \{128,256,512\}$ (mean of 100 runs), normalized to the latency of OS. Configurations on the y-axes are in the format of $(fw/fh,iw/ih,nf)$.}
\label{fig:bl-dfs}
\end{figure}

We compared the three basic dataflows (the experiment setup is outlined in Sec. \ref{sec:exp_setup}), and the results can be found in Fig. \ref{fig:bl-dfs}. We see that OS consistently outperforms the others in all tests conducted in terms of runtime. With a stride of 1, OS is by median 1.93x and 3.41x faster than IS and WS, respectively. With a stride of 2, OS is, by median, 5.39x and 2.81x faster than IS and WS, respectively. 
The superior performance of OS is due to a multitude of factors including lowered numbers of reduction sum operations, reduced output tensor data movement, and more regular instruction and memory access patterns. 

While basic dataflows capture the reuse opportunities of the data that are active in the current computation, they only utilize a limited number of vector registers (precisely $\frac{3 \times \text{vector variable size}}{\text{vector register size}}$), leaving all others idle. This is because, as discussed in Sec.~\ref{sec:intro}, compilers today are not able to discover vectorizable code - except for the simple cases - and fully utilize all vector registers automatically. This necessitates the need to extend the basic dataflows for faster inference.

\section{Extending the Basic Dataflows}\label{sec:ext-df}

We say that a dataflow utilizes the stationarity of some data if it keeps that data close to the compute units - in vector registers in our case - for reuse. A dataflow is $\sigma$ stationary if it uses $\sigma$ stationarity, where $\sigma$ is a predefined type of data (inputs, weights, or outputs).  We extend the notion of dataflow by defining two types of stationarities, i.e., \textit{anchoring stationarities} and \textit{auxiliary stationarities}.

\textit{Anchoring stationarity} is the stationarity that decides the execution order of computations. For example, output stationary dataflows have the outputs as their anchoring data type, so we always complete \textbf{all} computations involving an output element before moving on to the next. One dataflow can have at most one \textit{anchoring stationarity}. 
The most naive implementation of a dataflow is constituted of an anchoring stationarity only, which is equivalent to one of the basic dataflows discussed in Sec.~\ref{sec:bg-dataflows}. The major limitation of the basic dataflows is that not all vector registers are utilized.
    
In optimized implementations, vector registers are fully utilized to stash data to lower the data movement costs associated with both anchoring and non-anchoring data types -- non-anchoring data types are also referred to as \textit{auxiliary data types}. The \textit{auxiliary stationarities} determine which auxiliary data types should be allocated in vector registers. For example, an output-anchored dataflow may be accompanied by weight and/or input auxiliary stationarity. More than one auxiliary stationarity can accompany an anchoring stationarity. 
   
An important question is to decide how to allocate vector registers to store (or stash) anchoring and auxiliary data types, which is dependent on two factors: (1) the total number of available vector registers, which constraints the overall SIMD capability, and (2) data reuse opportunities, which affects data movement costs, and also bounds the benefits that can be obtained by stashing the corresponding data in vector registers. 

\section{Optimizing Extended Dataflows}\label{sec:motivation}
Our methodology for optimizing an extended dataflow follows two steps. First, we analyze reuse opportunities and develop heuristics to maximize data reuse benefits within each basic (i.e., anchoring stationarity only) dataflow to derive the corresponding auxiliary stationarities. Next, we empirically compare different implementations of the extended dataflows by varying vector register allocation schemes using a code generator to determine the best dataflow for performance. 

\begin{figure}
\includegraphics[width=\linewidth]{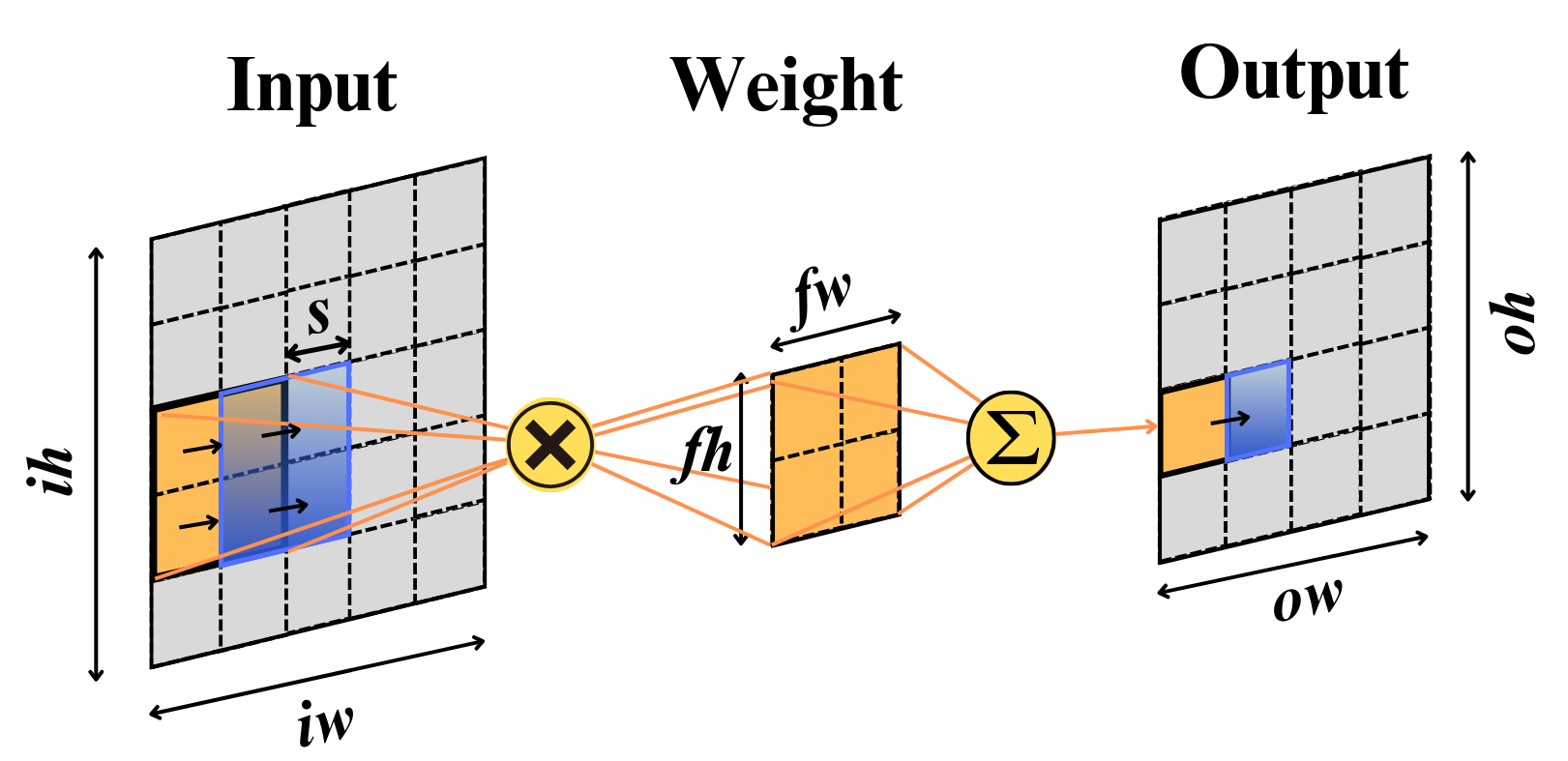}
\caption{Convolution operations and notations, showing only 1 channel and 1 kernel.}
\label{fig:conv_notation}
\end{figure}

While this methodology can be applied to most layers in neural networks, we focus our discussions on convolution layers, including simple convolutions \cite{lecun1995comparison}, depthwise convolutions \cite{sifre2014rigid, howard2017mobilenets}, grouped convolutions \cite{krizhevsky2012imagenet}, shuffled grouped convolutions \cite{zhang2017shufflenet}, and so on. This is because these layers are common, and their latencies are generally longer compared to other layers \cite{iandola2016squeezenet, sze2017efficient, howard2017mobilenets, chollet2017xception, zhang2018shufflenet}. 
The convolution operation is shown in Fig. \ref{fig:conv_notation}.
Notation-wise, we use $ih$, $iw$, $fh$, $fw$, $oh$, $ow$ for input height, input width, filter/weight height, filter/weight width, output height, and output width, $s$ for strides, $x$ for the number of data elements in a vector variable, and $H$, $R$, $E$ for the sizes of input, filter/weight, and output tensors. Thus, $H=ih\cdot iw\cdot x$, $R=fh \cdot fw \cdot x$, $E=oh \cdot ow$. 

\subsection{Maximizing Data Reuse under Each Basic Dataflow}
\subsubsection{Reuse under Output Stationary Dataflows}
\begin{figure}
    \centering
    \begin{subfigure}[b]{0.23\textwidth}
        \includegraphics[width=\textwidth]{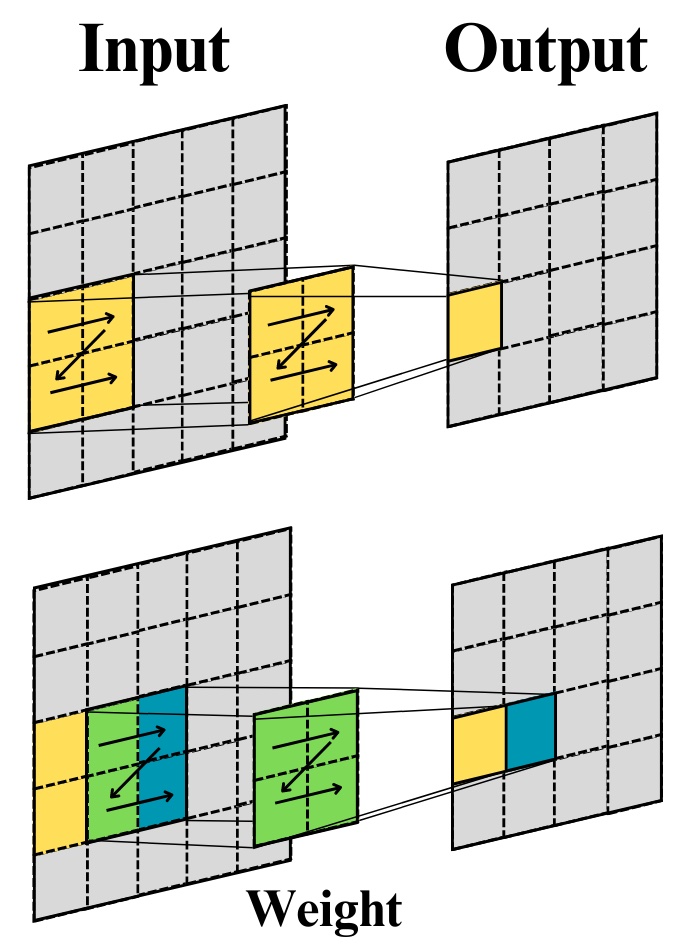}
        \caption{Output-anchored dataflows, stride 1, $2\times2$ weight filter.}
        \label{fig:reuse-os}
    \end{subfigure}
    \hfill
    \begin{subfigure}[b]{0.23\textwidth}   
        \centering 
       \includegraphics[width=\linewidth]{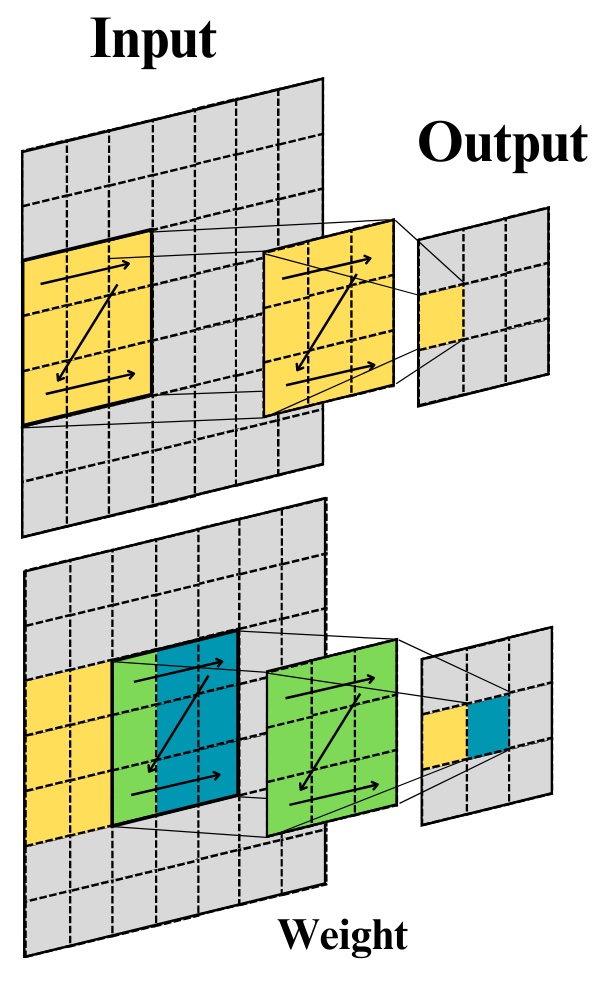}
        \caption{Output-anchored dataflows, stride 2, $3\times3$ weight filter.} 
        \label{fig:reuse-os-s2}
    \end{subfigure}
    
    \vskip\baselineskip
    \begin{subfigure}[b]{0.23\textwidth}   
        \includegraphics[width=\textwidth]{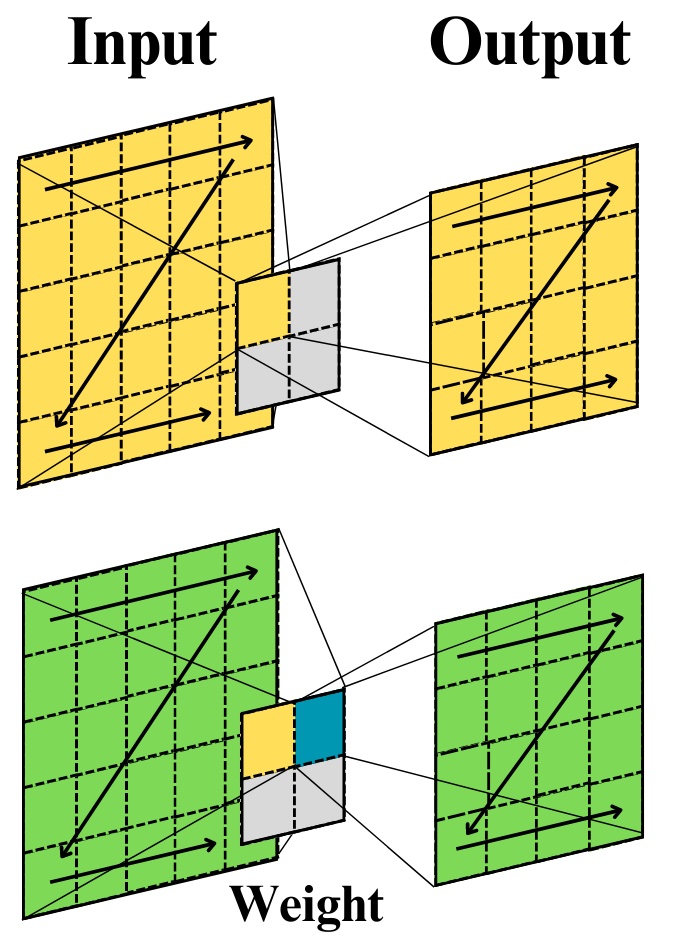}
        \caption{Weight-anchored dataflows, stride 1, $2\times2$ weight filter.}
         \label{fig:reuse-ws}
    \end{subfigure}
    \hfill
    \begin{subfigure}[b]{0.23\textwidth}  
        \includegraphics[width=\textwidth]{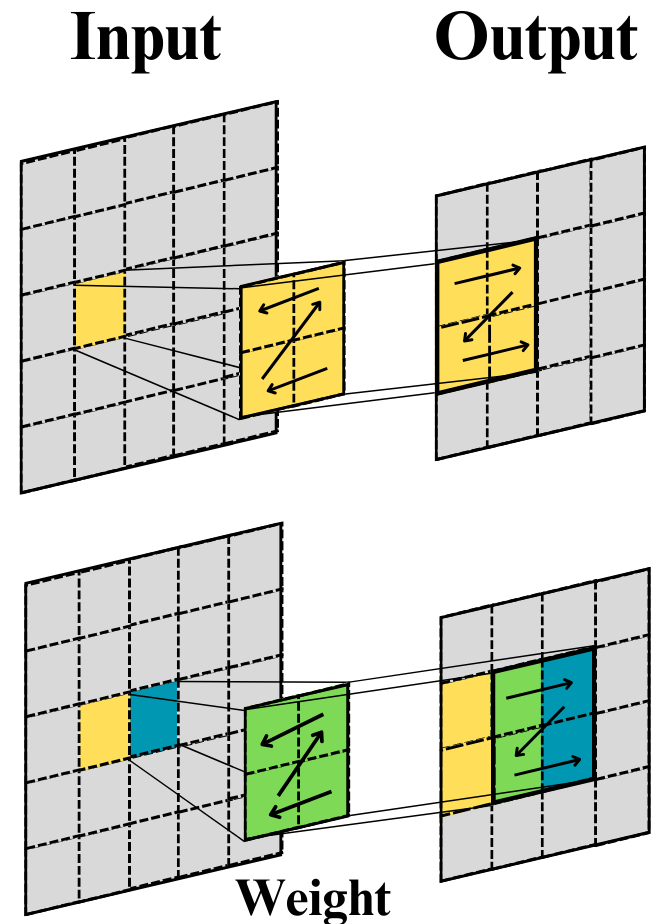}
        \caption{Input-anchored dataflows, stride 1, $2\times2$ weight filter.}
        \label{fig:reuse-is}
    \end{subfigure}
        \vskip\baselineskip
    \begin{subfigure}[b]{0.48\textwidth}   
        \centering 
       \includegraphics[width=\linewidth]{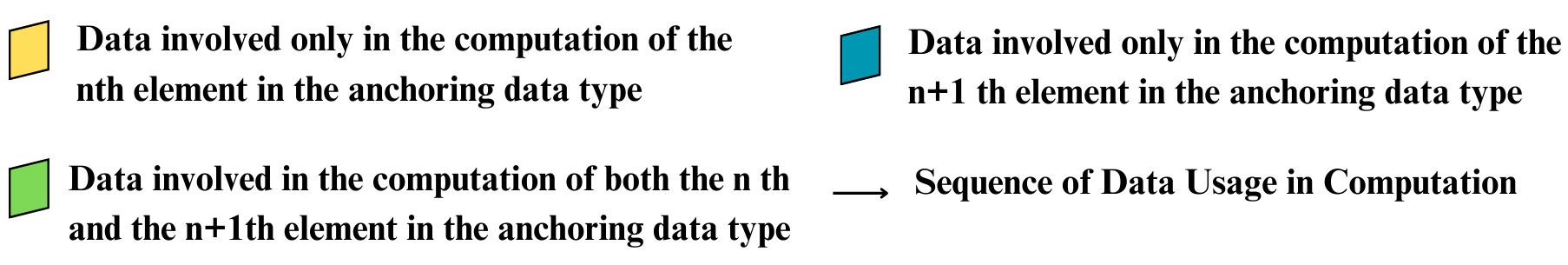}
        \label{fig:new_subfigure}
    \end{subfigure}
    \caption{Reuse opportunities under each anchoring dataflow, showing only one channel and one kernel.}
    \label{fig:reuse-all}
\end{figure}

Under output-anchored dataflows with the computation sequence following the description in Sec. \ref{sec:bg-dataflows-os}, all corresponding weights in each channel, totaling $R$, are reused between the computations for two successive output elements. Additionally, there are $(fw - s) \cdot fh$ reusable input elements involved in the computations for two successive outputs. We demonstrate these reuse opportunities in Fig. \ref{fig:reuse-os}.

The reuse scheme of inputs is similar for $s > 1$, as shown in  Fig. \ref{fig:reuse-os-s2}, differed only by the number of inputs reusable between the computations around two successive outputs.

\subsubsection{Reuse under Input Stationary Dataflows}

Given the algorithm of input-anchored dataflows (Sec. \ref{sec:bg-dataflows-is}), when $s = 1$, all corresponding weights in each channel, totaling $R$, can be reused between the computations around two successive input elements. Outputs (partial sums) under input-anchored dataflows can be reused in a way similar to how inputs are reused under output-anchored dataflows. We demonstrate this reuse scheme in Fig. \ref{fig:reuse-is}. Note that we would need to reverse the sequence of the weights (i.e., following the order of the outputs) to enable this reuse scheme (see Fig. \ref{fig:reuse-is}).

\begin{figure}
\includegraphics[width=\linewidth]{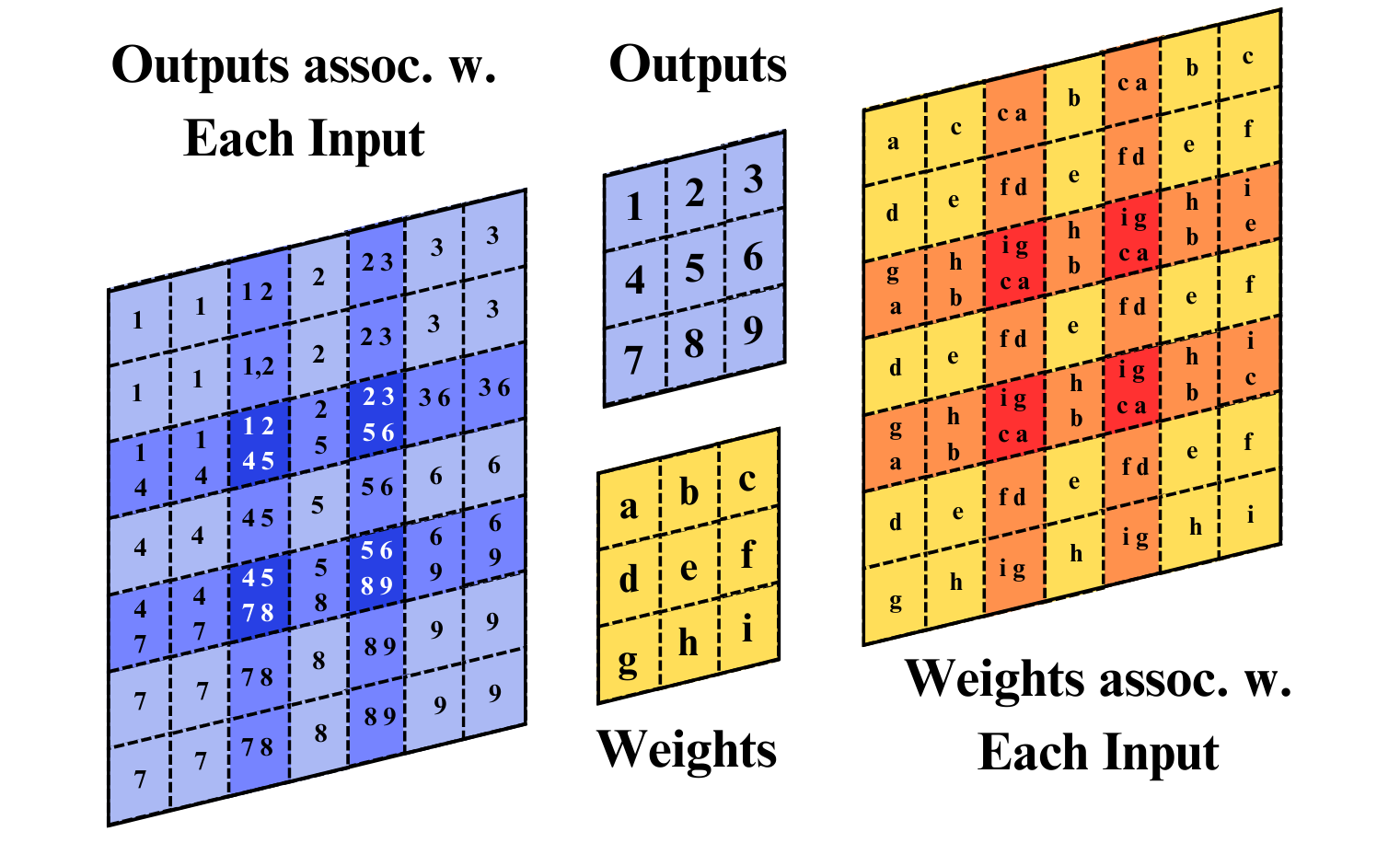}
\caption{Under input-anchored dataflows: weights and outputs associated with each input when s=2 for each channel. Darker color means more data are associated with that input element.}
\label{fig:reuse-is-s2}
\end{figure}

When $s>1$, reusing both outputs and weights becomes complicated. Not all weights are applied to every input. For $s=2$, the number of weights/outputs associated with the computations around one input can be 1, 2, or 4, as demonstrated in Fig. \ref{fig:reuse-is-s2}. In this case, the reuse opportunities become sparse. Additionally, code structure becomes less regular.

\begin{table*}[t]
    \centering
    \begin{threeparttable}
        \caption{Summary of gains from auxiliary allocation for each operation involving one channel block and one kernel}
        \label{tab:aux_gains}
        \begin{tabularx}{\textwidth}{c|c|c|c|X|X}
        \toprule
        Anc. Dataflow & Aux. & \# vector variables for aux. & Stride & Reduction in \# mem. reads for each additional vector variable allocated to auxiliary data & Reduction in \# mem. writes\tnote{a} for each additional vector variable allocated to auxiliary data \\
        \midrule
        OS & Both & $[1,R]$ & $[1,fw-1]$ & $E$ & 0\\
        \midrule
        \multirow{2}{*}{WS} & Input & $[1, H]$ & $[1, fw-1]$ & $R$ & 0 \\
        & Output & $[1, E]$ & $[1, fw-1]$ & $R$ & $R$\\
        \midrule
        \multirow{8}{*}{IS} & Weight & $[1,R]$ & 1 & $H$ & 0 \\
        & Weight & $[1,fw]$ & $[2,fw-1]$ & $\frac{H}{s}$ & 0\\
        & Weight & $[fw+1,2\cdot fw]$ & $[2,fw-1]$ & $\frac{H}{(fw-s)s}$ & 0\\
        & Output & $[1,R]$ & 1 & $H$ & $H$ \\
        & Output & \{1\} & $[2,fw-1]$ & $H+\frac{H}{fw}$ & $H+\frac{H}{fw}$\\
        & Output & \{2\} & $[2,fw-1]$ & $\frac{ih}{fw-s}(H+\frac{H}{fw})+\frac{ih}{s}(fw-s-1)$ & $\frac{ih}{fw-s}(H+\frac{H}{fw})+\frac{ih}{s}(fw-s-1)$\\
        & Output & $[3, (3+fw-s)]$ & $[2,fw-1]$ & $(fh-s)(fw-s)\frac{H}{R}$ & $(fh-s)(fw-s)\frac{H}{R}$\\
        \bottomrule
        \end{tabularx}
    \end{threeparttable}
\end{table*}

\subsubsection{Reuse under Weight Stationary Dataflows}\label{sec:reuse-ws}
In weight-anchored dataflows (Sec. \ref{sec:bg-dataflows-ws}), 
between the computations around two successive weights in an input channel block, all $H$ inputs and $E$ outputs can be reused, as depicted in Fig. \ref{fig:reuse-ws}.

When using vector registers to stash an input, the input will not be reused in the computation involving each weight when $s > 1$. On the other hand, stashed outputs are guaranteed to be reused with each weight. As stashing outputs also saves write-related operations and the size of the output tensor is almost always greater than the remaining SIMD vector registers, we will later demonstrate the sufficiency of only supporting output auxiliary stationarity under weight-anchored dataflows.

\subsubsection{Heuristics to Quantify the Effectiveness of Data Reuse under Each Dataflow} \label{sec:heuristics}
We use the reduction in the number of memory instructions (both read and write, data size = $c \times elem\_width$) for each input channel as the guiding metric for framing the heuristics for choosing auxiliary stationarities, summarized in Table \ref{tab:aux_gains}. 
The baseline configurations correspond to the basic dataflow implementations discussed in Sec.~\ref{sec:bg-dataflows}, where $3 \times \text{vector variable size} / {\text{vector register size}}$ vector registers are allocated only. For the extended dataflows, we utilize additional vector variables (which are mapped to vector registers) for the auxiliary data types to further reduce data movement costs.

\paragraph{Output-anchored Dataflows} Independent of the value of $s$, the numbers of inputs and weights associated with an output element, disregarding edge cases, are always equal to $R$ for each input channel. Thus, every time we stash an input or weight vector variable in one or more vector register(s), the number of memory reads always goes down by the size of the output tensor. 

\paragraph{Input-anchored Dataflows} When $s=1$, the gains from auxiliary allocation mimic that under output-anchored dataflows. We expect a reduction of $H$ memory reads and $H$ memory writes for every vector variable allocated to stash outputs for each input channel block. For each vector variable allocated for stashing weights, we expect a reduction of $H$ memory reads per input channel block. Note that $H \approx E$ in this case.
When $s > 1$, the gains from auxiliary allocation become complex as shown in Table~\ref{tab:aux_gains}. 

\paragraph{Weight-anchored Dataflows} Recall from Sec. \ref{sec:reuse-ws} that we iterate through both the whole input and output tensors under weight-anchored dataflows. While we proceed by $1$ element on the output tensor, we need to leap forward by $s$ elements on the input tensor and also increment the starting input index (i.e., the first weight starts with the input at index $0$, the second weight starts with the input at index $1$, and so forth) for the computations associated with each weight element. This naturally implies that each vector variable allocated for inputs saves $R \approx \frac{H}{s^{2}}$ memory reads, and each vector variable assigned to stash outputs saves $R$ reads and $R$ writes, respectively, per input channel block.

\noindent{}Guided by the heuristics, we derive the following observations:

\textbf{Observation 1:} Weight-anchored dataflows will gain the least performance improvement from auxiliary stationarities.

\textbf{Observation 2:} Output-anchored dataflows will likely yield better performance than input-anchored dataflows when both are fully optimized.

\textbf{Observation 3:} Under output-anchored dataflows, prioritizing input auxiliary stationarity and prioritizing weight auxiliary stationarity will yield similar results.

\textbf{Observation 4:} Under input-anchored dataflows, prioritizing output auxiliary stationarity will yield better performance than prioritizing weight auxiliary stationarity.

\textbf{Observation 5:} Under weight-anchored dataflows, prioritizing output auxiliary stationary will yield better performance than prioritizing input auxiliary stationary.
        
\subsection{Extended Dataflow Implementations and Code Generator}
Based upon the above observations, we develop a code generator to extend all three basic anchoring dataflows with auxiliary stationarities to further determine vector register allocation schemes, which is done by varying the number of vector registers allocated to each type of data. We first allocate a subset of vector registers (sweeping from $v_0$ to $v_{3n-1}$, where $n = size(vec\_var)/size(vec\_reg), size(vec\_var) \in \{128,256,512\}, \text{and } size(vec\_reg)=128$ in our implementation) to store the vector variables corresponding to the anchoring data type, then the remaining vector registers to the auxiliary data types. 
\begin{algorithm}
\small
\caption{Allocation sequence for inputs under secondary-unrolled output-anchored dataflows. (The same sequence applies for outputs under input-anchored dataflows when $s = 1$.)}
\label{alloc_sequence}
\begin{algorithmic}
\State Initialize the original allocation sequence with sequential row-major allocation.
\For{$un$ in range[1, $lcm$(all \#vector variables per row $> stride$))}
    \If{\# vector variables on this row $> stride$}
    \State Rotate stash indices on this row left by $stride$
    \Else \State The sequence stays the same
    \EndIf
\EndFor
\end{algorithmic}
\end{algorithm}

\subsubsection{Implementation of Output-anchored Dataflows} \label{sec:imp-os}
For each output element under computation, we first determine if the required input and weight elements are already stashed in vector variables. If so, we perform the computation using those stashed data. Otherwise, we load the required data from memory into 2 vector variables of length $size(vec\_var)=x\times element\_width$. Note that the sequence of vector variable usage between every two consecutive outputs is identical for weights but different for inputs. This means that we incur the cost of SIMD data transfer if we assign vector registers in the same way across all unrolled iterations of the weight loop, as the same position on the ``window" covering all inputs involved in the computations of an output data would be matched to a different input in two successive iterations.

\begin{figure}
\includegraphics[width=\linewidth]{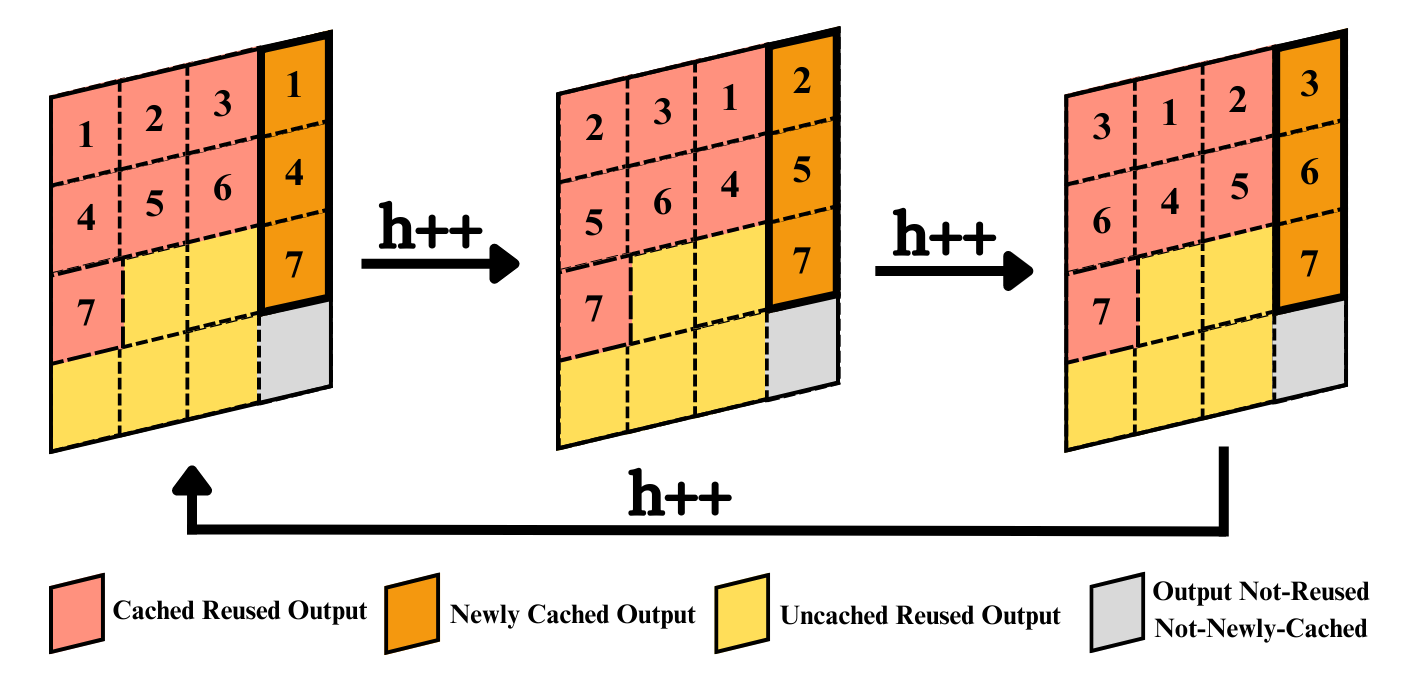}
\caption{Secondary loop unrolling to bypass vector data transfer, using one channel for demonstration.s}
\label{fig:second-unroll}
\end{figure}

To circumvent unnecessary data transfers between vector registers used for auxiliary input stationarity, we implement \textit{secondary unrolling}, performed on the output loop with a magnitude of the least common multiple of all numbers of input vector variables per row (in the input tensor) that are greater than $s$, so that each iteration of the secondary unrolled loop uses vector variables differently: the specific sequence of allocating input vector variables differs between the computations around two successive outputs if the number of input vector variables in that row is greater than $s$, and remains the same otherwise.
Algorithm \ref{alloc_sequence} demonstrates the sequences of vector variable allocations for input auxiliary stationarity across each secondary-unrolled iteration, and Fig. \ref{fig:second-unroll} provides a graphical example of secondary loop unrolling. 

To further minimize data movements, we directly load vectors of input data to be newly stashed into their corresponding vector variables (thereby overwriting the previous data), instead of new vector variables.

It is also worth noting that, 
through our observations, we found it advantageous to accumulate all results in a single vector register (instead of a scalar register) and execute the reduction sum operation only when all computations involving an output element have been completed. Although this approach consumes more vector registers, it ultimately saves costs related to performing a reduction sum operation on a scalar variable upon the completion of each computation. 

Algorithm \ref{alg:os-opt} summarizes the implementation of output-anchored dataflows.

\begin{algorithm}
\caption{Implementation of Output-anchored Dataflows}
\footnotesize
\label{alg:os-opt}
\begin{algorithmic}
\Require $inputs[ih\cdot iw \cdot ic]$, $weights[ih\cdot iw \cdot ic \cdot oc]$, $outputs[oh \cdot ow \cdot ic \cdot oc]$
\Require $numInStash$, $numWgtStash$, $x$, $s$
    \State \textbf{Prep 1:} Initialize a total of $numInStash$ input vector variables by loading data from the input tensor.
    \State \textbf{Prep 2:} Initialize a total of $numWgtStash$ weight vector variables by loading data from the weight tensor.
    \State
    \For{$c$ in $ic$ by $x$}
        \For{$k$ in $oc$}
            \For{$h$ in $oh$ by $s$}
                \For{$w$ in $ow$ by $s$} \Comment{Secondary Unroll}
                    \State Set the anchoring $output$ vector variable to $\Vec{0}$
                    \For{$r$ in $fh$} \Comment{Unroll}
                        \For{$s$ in $fw$} \Comment{Unroll}
                            \If{$r\cdot fh+fw < numInStash$}
                                \State Use the stashed vector as input
                            \ElsIf{$r\cdot fh + (fw - s) < numInStash$}
                                \State Overwrite a completely used input stash with the new input by $vload(c\cdot ih\cdot iw + h \cdot iw + w)$ and then use it as $input$
                            \Else
                                \State $input = vload(c\cdot ih\cdot iw + h \cdot iw + w)$
                            \EndIf
                            \If{$r\cdot fh + fw < nunWgtStash$}
                                \State Use the stashed vector as weight
                            \Else
                                \State $weight = vload(c\cdot oc \cdot ih \cdot iw+ k\cdot fh \cdot fw +r \cdot fw +s)$
                            \EndIf
                            \State$res=vmul(input,weight)$
                            \State$output=vadd(output,res)$
                        \EndFor
                    \EndFor
                    \State$output[k\cdot oh \cdot ow + h \cdot ow + w] += vaddv(output)$
                \EndFor
            \EndFor
        \EndFor
    \EndFor
\end{algorithmic}
\end{algorithm}

\begin{algorithm}
\caption{Implementation of Input-anchored Dataflows}
\small
\label{alg:is-opt}
\begin{algorithmic}
\Require $inputs[ih\cdot iw \cdot ic]$, $weights[ih\cdot iw \cdot ic \cdot oc]$, $outputs[oh \cdot ow \cdot ic \cdot oc]$
\Require $numInStash$, $numWgtStash, x, s$
    \State \textbf{Prep 1:} Initialize a total of $numInStash$ input vector variables by loading data from the input tensor.
    \State \textbf{Prep 2:} Initialize a total of $numWgtStash$ weight vector variables by loading data from the weight tensor.
    \State

    \For{$c$ in $ic$ by $x$}
        \For{$k$ in $oc$}
            \For{$h$ in $ih$}
                \For{$w$ in $iw$} \Comment{Secondary Unroll}
                    \State $input = vload(c\cdot ih\cdot iw + h \cdot iw + w)$
                    \For{$((h',w'),(r,s))$ in $(assoc\_idx(h,w,c))$}  
                        \State\Comment{In reverse order}
                        \State \Comment{Output and weight indices. See Fig. \ref{fig:reuse-is-s2}}
                        \If{$r \cdot fw + s \in stashedWeightsIndices$}
                            \State Use stashed vector as weight
                        \Else
                            \State $weight = vload(c\cdot oc \cdot ih \cdot iw+ k\cdot fh \cdot fw +r \cdot fw +s)$
                        \EndIf
                        \If{$h' \cdot iw + w' \in stashedOutputIndices$}
                            \State Use the stashed vector as output
                            \State $res = vmul(input,weight)$
                            \State $output = vadd(res,output)$
                            \If{Last use of this output}
                                \State$output[k\cdot oh \cdot ow + h \cdot ow + w] += vaddv(output)$
                            \EndIf
                        \ElsIf{$h' \cdot iw + w'$ to be newly stashed}
                            \State Use a free vector as output
                            \State $output = vmul(input,weight)$
                        \Else
                            \State $output[k\cdot oh \cdot ow + h \cdot ow + w] += vaddv(vmul(input,weight))$
                        \EndIf
                    \EndFor
                    \State$output[k\cdot oh \cdot ow + h \cdot ow + w] += vaddv(output)$
                \EndFor
            \EndFor
        \EndFor
    \EndFor
\end{algorithmic}

\end{algorithm}

\subsubsection{Implementation of Input-anchored Dataflows}
Under input-anchored dataflows, we can allocate the remaining vector variables to both weights and outputs. 
When $s$ is 1, we observe that the sequences of vector variable usage between every two consecutive inputs are identical for weight data but different for output data. Similar to the output-anchoring dataflows, this means that we incur the cost of vector data transfer if we consistently use variables in the same sequence. Therefore, again, we perform secondary unrolling on the output loop, following a similar procedure as described in Sec. \ref{sec:imp-os}, but with the sequence of weights in reverse. 
We write the stashed outputs back to memory when their usage is complete for this row, i.e., when the output is in the first column of the current window of computation. The pseudocode of Input-anchored dataflows is provided in Algorithm \ref{alg:is-opt}.

\begin{algorithm}
\caption{Implementation of Weight-anchored Dataflows}
\label{alg:ws-opt}
\small
\begin{algorithmic}
    \Require $inputs[ih\cdot iw \cdot ic]$, $weights[ih\cdot iw \cdot ic \cdot oc]$, $outputs[oh \cdot ow \cdot ic \cdot oc]$
    \Require $numInStash$, $numOutStash$s
    \State \textbf{Prep 1:} Initialize a total of $numInStash$ input vector variables by loading data from the $input$ tensor.
    \State \textbf{Prep 2:} Initialize a total of $numOutStash$ output vector variables by setting them to $0$'s.
    \State

    \For{$c$ in $ic$ by $x$}
        \For{$k$ in $oc$}
            \For{$fi$ in $fh\cdot fw$} \Comment{Unroll to Split the last iteration}
                    \State $weight = vload(c\cdot oc \cdot ih \cdot iw+ k\cdot fh \cdot fw + fi)$
                        \For{$h$ in $oh$}
                            \For{$w$ in $ow$} 
                                \State Calculate $ih$ and $iw$ with $oh$, $ow$, $padding$, $s$
                                \If{$ih \cdot input\_width + iw  < numInStash$}
                                    \State Use the stashed vector as input
                                \Else
                                    \State $input = vload(c\cdot ih\cdot iw + h \cdot iw + w)$
                                \EndIf
                                \If{$h \cdot ow + w < numOutStash$}
                                    \State Use the stashed vector as output
                                    \State $output = vadd(vmul(input,weight))$
                                    \If{$fi == fh\cdot fw -1$}
                                        \State $outputs[k\cdot oh \cdot ow + h \cdot ow + w] += vaddv(output)$
                                    \EndIf
                                \Else
                                    \State $outputs[k\cdot oh \cdot ow + h \cdot ow + w] += vaddv(vmul(input,weight))$
                                \EndIf
                            \EndFor
                        \EndFor
            \EndFor
        \EndFor
    \EndFor
\end{algorithmic}
\end{algorithm}

\subsubsection{Implementation of Weight-anchored Dataflows}
Similar to output- and input-anchored dataflows, we describe a concrete and general method to implement weight-anchored dataflows in Algorithm \ref{alg:ws-opt}. For input and output auxiliary stationarity under weight-anchored dataflows, we always stash the earliest yet unstashed element to exploit locality. We perform a loop split on the weight loop on top of unrolling to write stashed outputs back to memory only when their last usage is complete. When $s>1$, inputs are reused once for every $s$ weights. 

Our code generator follows Algorithms \ref{alg:os-opt}, \ref{alg:is-opt}, and \ref{alg:ws-opt} to implement various extended dataflows using ARM Intrinsics. 
Users input the anchoring stationarity, the number of vector variables to be allocated to each auxiliary stationarity, and the layer configurations to generate custom dataflow implementations.

\subsection{End-to-End Optimization of Memory Layout Sequence}\label{sec:ext_dfs:end_to_end}

Consistent memory layout alignment across consecutive layers is a prerequisite for efficient neural network inference. Any layout discrepancy entails the need for transformations, leading to additional overhead. To combat this issue, we resort to the commonly adopted dynamic programming approach based on searched results \cite{bib:liuatc, ahn2020ordering,lu2018spwa}. The algorithm's strategy hinges on minimizing layout transformations by using costs obtained from repeated runs of different scheduling schemes on each layer, ensuring reduced variance. By leveraging these costs, the algorithm determines optimal layouts that synchronize every two successive layers, thus curtailing the necessity for layout transformations.

In addition, we search for the optimal blocking schemes in compile time by running the program under each of the possible configurations and comparing their performance.







\section{Experiment Setup} \label{sec:exp_setup}
We use physical ARM machines to quantitatively evaluate and compare dataflows implemented using our code generator. These experiments encompass executing convolution layers with various combinations of the following parameters, as well as collecting end-to-end runtime results for neural networks, to facilitate a thorough and comprehensive evaluation and comparison of different dataflows.

\begin{itemize}
\item \textbf{Input Size:} We focus on larger convolution layers that are time-consuming with input sizes of $56\times56$ and $112\times112$.
\item \textbf{Weight filter Size:} We use filters of sizes $3\times3$, $4\times4$, and $5\times5$, as these dimensions are most widely employed.
\item \textbf{Stride:} We use strides of $1$ and $2$, as these values are also the most commonly used.
\item \textbf{Number of Filters:} We tested with $128$, $256$, and $512$ filters to compare the different dataflows across various numbers of filters.
\item \textbf{Vector Lengths:} $128$, $256$, and $512$, which are supported by modern ISAs such as ARM\cite{arm2023neon} and x86 \cite{intel2023amx,intel2023intrinsics}.
\end{itemize}

We use the GCC compiler \cite{gcc} with the most aggressive optimization flags to compile all programs. We ran our experiments on a system with 64-bit quad-core ARM Neoverse-N1 CPUs which adopts the aarch64 architecture. Each program was executed 100 times to obtain the average run time.

\section{Results and Discussions}
\subsection{Validation of Heuristics} \label{sec:exp1}
We generated programs that implement extended dataflows for various convolution layers in ARM Intrinsics and ran experiments following the setup described in Sec. \ref{sec:exp_setup} to validate the heuristics described in Sec. \ref{sec:motivation}.

We primarily present the results for $s=1$ because
(1) With output-anchored dataflows, the relative gains from weight and input auxiliary stationarities stay constant regardless of whether $s$ is $1$ or $2$. 
(2) For weight-anchored dataflows, according to our heuristics in Sec. \ref{sec:heuristics}, the improvement of extended dataflows over the basic, anchoring-only dataflow under $s=2$ is expected to be less than that for $s=1$. 
(3) Under input-anchored dataflows, as $s$ increases, the difference between gains from weight and output auxiliary stationarity amplifies -- we have empirically observed this behavior. 
(4) To compare output-anchored and input-anchored dataflows, we aim to determine whether the additional memory writes due to auxiliary output stationarity can exceed the 1.93x difference. Studying this under $s=2$ is less insightful, as the difference between OS and IS (5.39x) is considerably larger. 

\subsubsection{Comparing Different Anchoring Stationarities}

\begin{mybox}
\begin{finding}
Weight-anchored dataflows yield the least improvement from auxiliary dataflow optimizations and are consistently the slowest by a large magnitude. 
\end{finding}
\end{mybox}

\begin{figure}[!htb]
  \centering
  
  \begin{subfigure}[b]{\linewidth}
    \includegraphics[width=\linewidth]{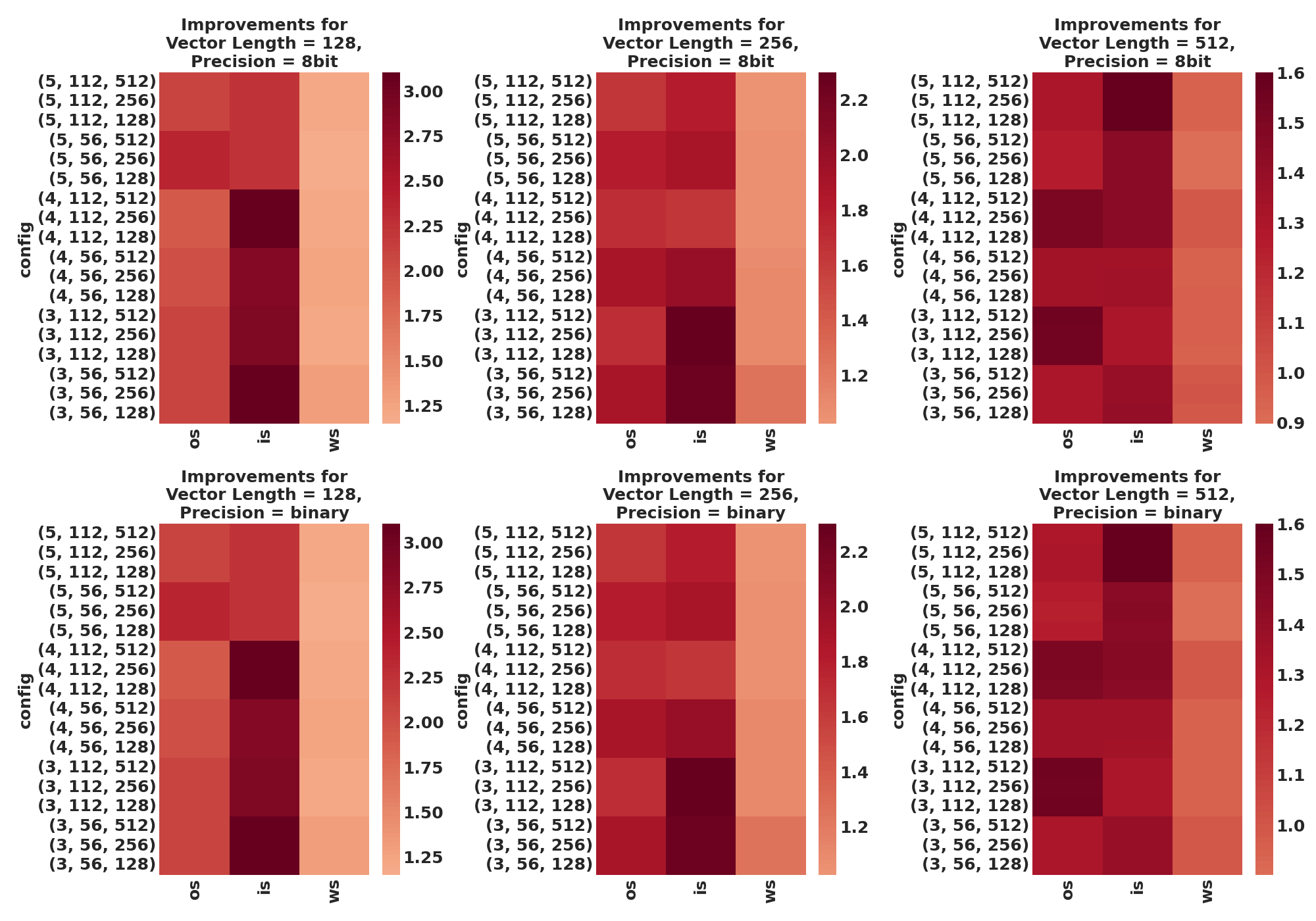}
    \caption{Speedup from the most optimized extending dataflows, normalized to the respective basic dataflows (i.e., results from Fig.~\ref{fig:bl-dfs}).}
    \label{subfig:ext_bl_comp}
  \end{subfigure}
  
  \vspace{0.2cm}  
  
  \begin{subfigure}[b]{\linewidth}
    \includegraphics[width=\linewidth]{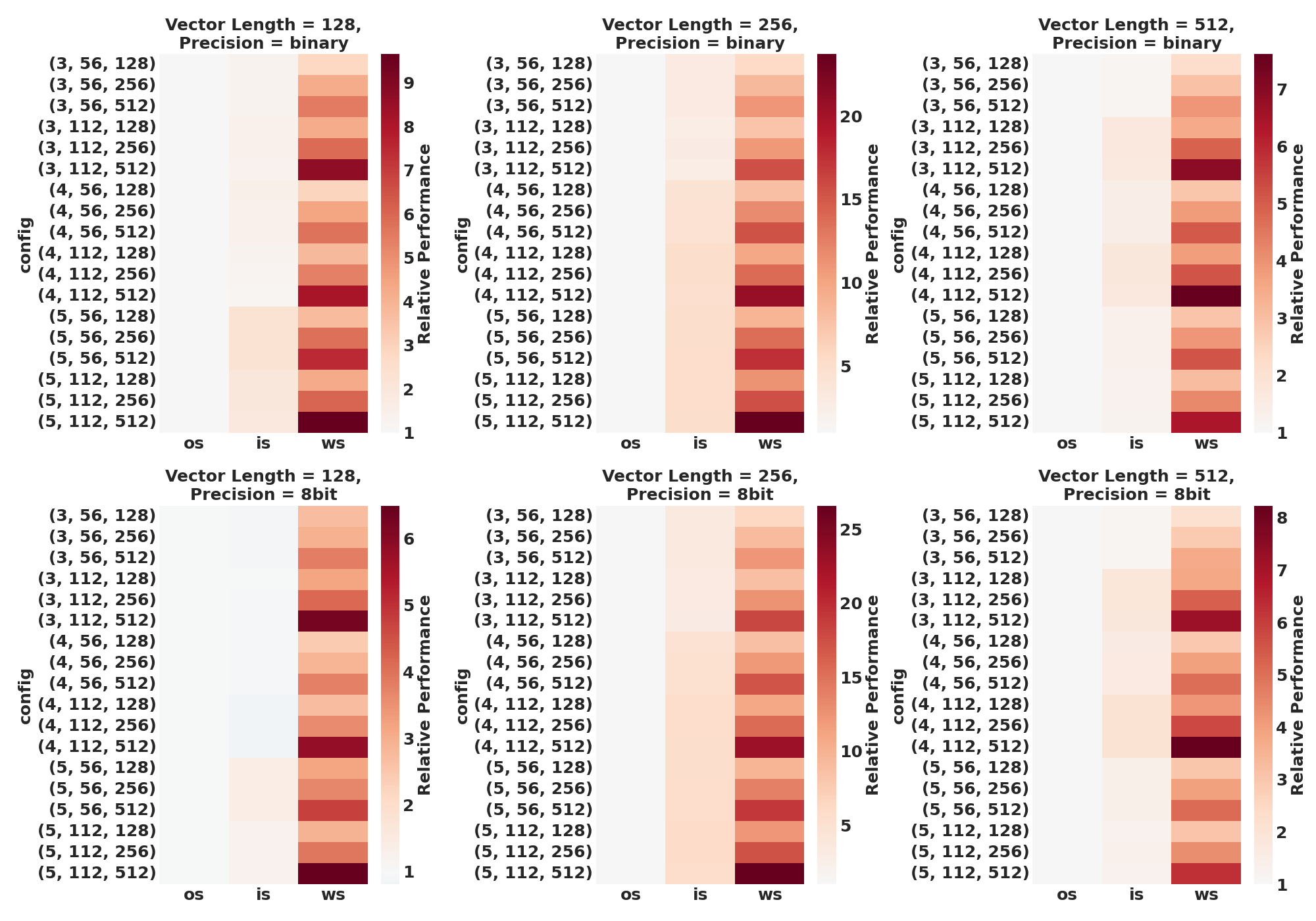}
    \caption{Relative Latency comparing the most optimized extended dataflows, normalized to the performance of OS.}
    \label{subfig:ext_latency}
  \end{subfigure}
  
  \caption{Performance results of extended dataflows. $\text{Vector Length} = (elem\_width \times c) \in \{128,256,512\}$ (mean of 100 runs). Configurations on the y-axes are in the format of $(fw/fh,iw/ih,nf)$.}
  \label{fig:combined_figure}
\end{figure}

Weight-anchored dataflows, even when fully optimized, significantly underperform in comparison to other anchoring stationarities (Fig. \ref{subfig:ext_latency}). Surprisingly, fully optimized output-anchored dataflow implementations are by median approximately 7.41x faster than their weight-anchored counterparts. However, when comparing the basic dataflows, we observe only a median performance difference of about 5.44 times between WS and OS, and roughly 2.91 times between WS and IS, given $s=1$. This escalating disparity is attributed to the different performance enhancements yielded by our optimization technique for different anchoring dataflows. As illustrated in Fig. \ref{subfig:ext_bl_comp}, the introduction of auxiliary stationarities results in a modest median improvement of around x1.08 for WS, while IS and OS enjoy more substantial median speedups of approximately x1.96 and x1.78 times, respectively. In fact, we find that adding auxiliary stationarities to the basic WS dataflow can sometimes lengthen the compute time. This is due to a low reuse frequency of the stashed auxiliary data and a more dominant increase in the size of the instruction cache. This result validates \textbf{Observation 1} derived from our heuristics.
\begin{mybox}
\begin{finding}
Output-anchored Dataflows outperform input-anchored Dataflows in the majority of the cases.
\end{finding}
\end{mybox}
While IS seems to gain a larger performance improvement from the addition of auxiliary stationarities, we still find output-anchored dataflows to be superior upon full optimization. For the same convolution layer configuration, optimized output-anchored dataflows are faster than input-anchored dataflows for around 90\% of the cases, which validates \textbf{Observation 2}.

\subsubsection{Findings Related to Auxiliary Stationarity}
Here, we compare different auxiliary stationarity schemes under each anchoring dataflow.

\begin{mybox}
\begin{finding}
Prioritizing stashing inputs or weights does not significantly impact performance under output-anchored dataflows.
\end{finding}
\end{mybox}

This finding validates \textbf{Observation 3}. By comparing the latency of dataflows that prioritize allocation for weight auxiliary stationarity and the ones that prioritize input auxiliary stationarity, we observe neither allocation scheme is consistently superior to the other, and the differences between the two schemes are small (within 6\%). 

\begin{mybox}
\begin{finding}
Allocating vector variables to outputs first improves performance compared to prioritizing allocation for weights under input-anchored dataflows.
\end{finding}
\end{mybox}

By average, prioritizing stashing outputs yields an 8\% performance gain, which becomes more evident as we increase the vector length. It follows that \textbf{Observation 4} is validated.

\begin{mybox}
\begin{finding}
Prioritizing output allocation yields only slightly better performance than prioritizing input allocation under weight-anchored dataflows.
\end{finding}
\end{mybox}\label{sec:ws-verify}

We find that under almost all cases, prioritizing output auxiliary stationarity brings a performance gain of up to 3\% over prioritizing weight auxiliary stationarity. This validates \textbf{Observation 5}; however, the differences are negligible. 

\begin{algorithm}
\caption{Optimized Dataflow: Output Anchored Stationarity with Weight Auxiliary Stationarity}
\small
\begin{algorithmic}
\Require $numVecReg$, $vecVarSize$, v$ecRegSize$
\State $regsPerVar$ = $vecVarSize$ / $vecRegSize$
\State $numVarAvailable$ = $numVecReg$ / $regsPerVar$
\State
\State $auxVarAvailable$ = $numVarAvailable - 3$
\State
\State 1. Use \textbf{output stationary} as the anchoring stationarity
\State 2. Allocate $auxVarAvailable$ vector variables \textbf{first to weight}
\State \textbf{and then to input} (if there are still some remaining).
\end{algorithmic}
\label{alg:final}
\end{algorithm}

\subsubsection{Optimized Dataflow}
From all previous analyses and results, we conclude that OS-anchored dataflow with auxiliary weight stationarity is the most optimized dataflow in our study. While there is generally little difference between prioritizing auxiliary WS and prioritizing auxiliary IS, we find the former to yield better code readability and more regular instruction patterns. Algorithm \ref{alg:final} summarizes this dataflow.

\subsection{Neural Network Speedup against State-of-the-Art Implementations} \label{sec:result:speedup}

Applying end-to-end optimizations discussed in Sec. \ref{sec:ext_dfs:end_to_end}, we compare our technique to state-of-the-art baselines. 

For INT8 neural networks, we use TVM as one of the baselines. TVM is a highly optimized machine learning compiler stack for efficient neural network deployment across various hardware platforms \cite{bib:tvm}. We compare the end-to-end inference latency of variants of ResNet \cite{he2016deep} (Resnet-18 and Resnet-34) and VGG \cite{simonyan2014very} (VGG-11, VGG-13, and VGG-16) with TVM-autotuned (we use GridSearchTuner as the KernelTuner - this enumerates through the entire search space for configurations \cite{autotvm}) implementations and untuned implementations (TVM default). We set TVM to target the architecture and SIMD extension to match the physical machines used for our experiments. Across all network architectures and numbers of threads, we observe a $\sim$3x speedup over TVM's implementations, and up to $\sim$14x over its untuned implementation. Moreover, our multithreading scheme yields comparable scalability. We also compare the end-to-end results with programs generated by gcc/clang (with the highest level of optimization and autovectorization enabled). Ours achieve significant (4x-6x) speedup.

For the evaluation of binary neural networks, we compared the inference latency of our implementations with Cowan et al.'s TVM-based bitserial implementations \cite{cowan20automatic}. 
Since the code released by Cowan et al. only works for convolution layers on CPUs (while their end-to-end code generation tool targets Raspberry Pi and is not applicable to CPUs), we only perform this comparison for convolution layers. 
Bitserial implementations, although optimized for low-power consumption, do not offer satisfactory inference speed. Notably, our implementations are over 12x faster for various convolution layers.
Based on the end-to-end results reported in their paper  (which incorporates additional optimizations through microkernel synthesis) \cite{cowan20automatic}, we anticipate that our implementations will still outperform theirs by a large margin (6x or higher) in the end-to-end comparisons. 
We also compared our implementations of various convolution layers in VGG against those from \cite{bib:liuatc},  and ours achieve up to 4.8x speedup.




\begin{figure} 
  \includegraphics[width=\linewidth]{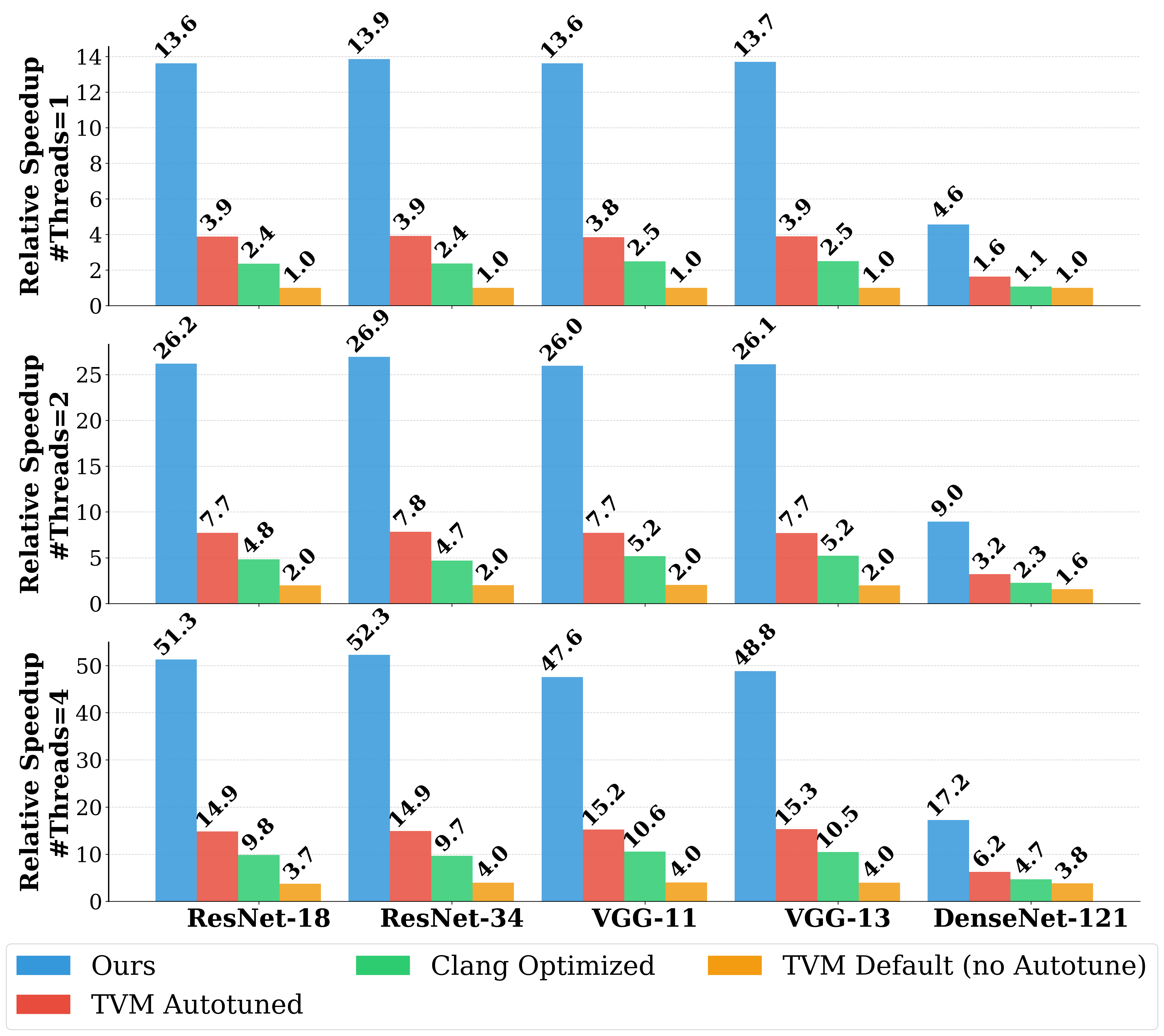}
  \caption{End-to-end relative speedups for 8-bit quantized neural networks from our techniques, normalized to TVM default mode without autotune (Note: for DenseNet-121 we do not have the results for TVM default mode, and had to use a different tuner (TaskScheduler), and we use the first tuning trial as the baseline).}
  \label{fig:end_to_end}
\end{figure}

\begin{figure} 
  \includegraphics[width=\linewidth]{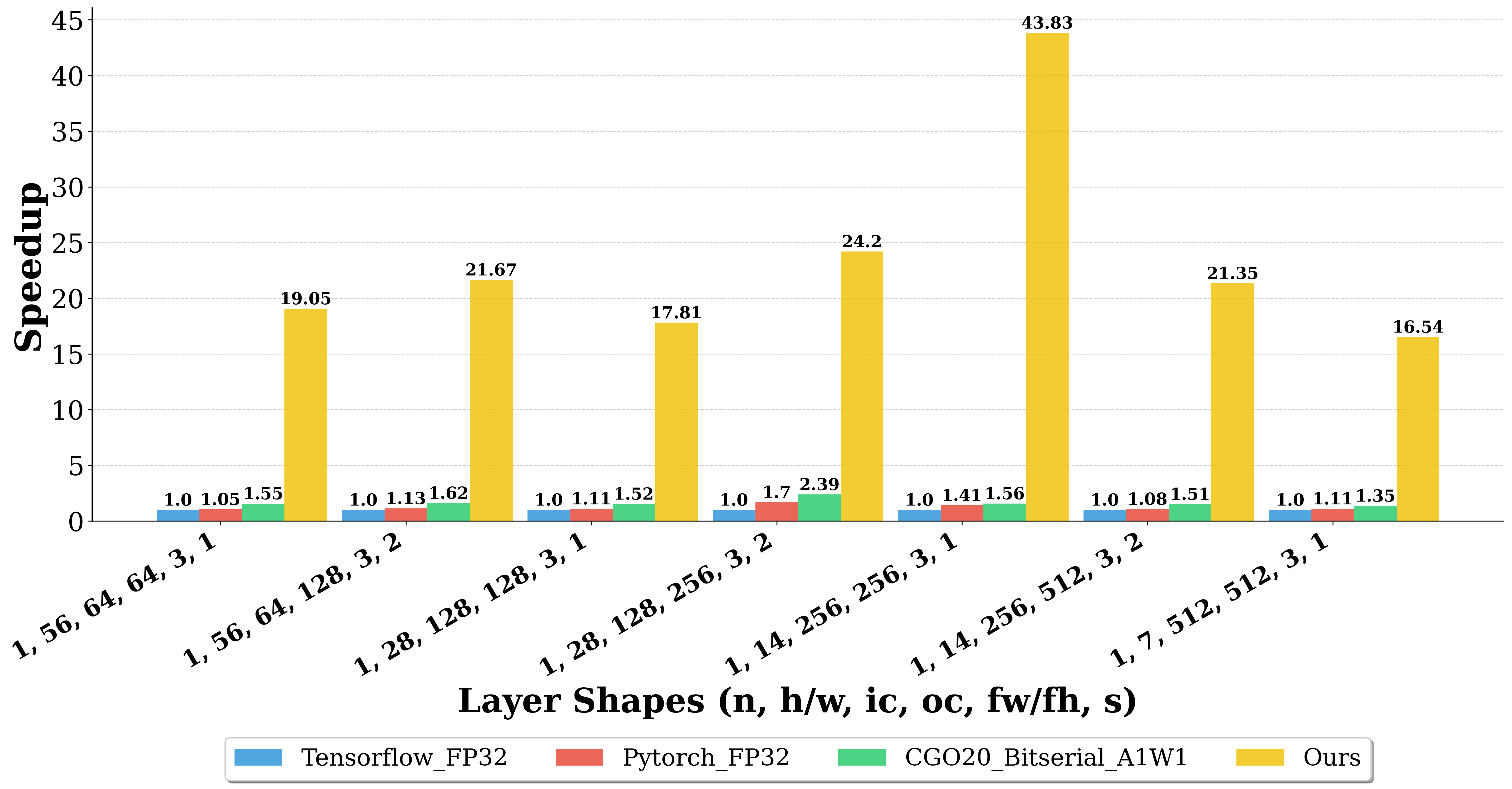}
  \caption{Layer-wise latency comparisons for binary Resnet Workloads between Ours and Cowan et al \cite{cowan20automatic}.}
  \label{fig:layerwise_binary}
\end{figure}

\section{Related Work}\label{sec:other_techniques}
This section offers an overview of prevalent techniques for accelerating neural network inference. Our work already employs quantization \cite{han2016deep, zhou2016dorefa, courbariaux2015binaryconnect, hoffer2017train, choi2018towards}, vectorization \cite{bib:halide,bib:interstellar,bib:depthconv,bib:directconv,bib:bitflow}, tiling/blocking \cite{bib:purireorder,bib:vastensor}, operator fusion \cite{bib:rhumem,bib:jiafusion,bib:qiaofusion}. We compare and contrast our work with other related efforts.

\paragraph{Unroll-and-Jam}\label{sec:unroll_and_jam}
Unroll-and-jam 
reduces memory access costs by reordering instructions without breaking data dependencies \cite{carr1996improving, mellor2004optimizing, carr1997unroll,stock2012using}, which can enhance the performance of convolution and fully-connected layers in DNNs \cite{bib:tvm, mandal2017optimizing, venkat2019swirl}. Our technique bypasses unneeded load instructions previously handled by jamming, and further jamming can be applied on top of our technique to lower latency.

\paragraph{Winograd Convolution}
Winograd convolution reduces the complexity of convolution operations \cite{winograd1980arithmetic,liu2018efficient,meng2019efficient,yan2020optimizing,alam2022winograd} and there exist various optimizations of its implementation on CPUs \cite{zlateski2019anatomy, jia2018optimizing, li2021lowino, maji2019efficient, li2021optimizing}. Utilizing a similar concept of reusing data to speed up convolution inference, DREW \cite{DREW} optimizes Winograd convolution by clustering data and reusing computed results and trades off accuracy and inference performance. In contrast, our method retains accuracy and suits all architectures with SIMD support. Moreover, standard Winograd convolutions struggle with quantization \cite{li2021lowino, chikin2022channel, fernandez2020searching, Fernandez-Marques2020}, while our technique does not suffer from this limitation.

\paragraph{Transformer Optimizations}
Transformers have revolutionized several areas of machine learning \cite{khan2022transformers, vaswani2017attention, subakan2021attention, shamshad2023transformers, yang2020clinical, yang2021transformer}. However, optimizing their performance, particularly on CPUs, remains a significant challenge \cite{ivanov2021data, wang2020hat, jiang2022characterizing, dice2021optimizing}. Efforts to date include pruning \cite{lagunas2021block, kwon2022fast, zhu2021vision, mao2021tprune}, quantization \cite{liu2021post,bondarenko2021understanding, chung2020extremely, prato2019fully}, knowledge distillation \cite{chen2022dearkd, jiang2021knowledge,liu2022transkd, wang2020minilm}, architecture search \cite{wang2020hat,liu2022uninet,yin2021autotinybert}, GEMM optimizations \cite{dice2021optimizing,jiang2022characterizing,hong2022dfx}, and hardware-level optimizations \cite{ivanov2021data, zhang2023niot}. 
Moreover, while there exist previous works on studying dataflows for transformers on other hardware platforms \cite{you2023vitcod, lu2021sanger, shen2022salo, zhao2022fpga}, no dataflow work has been done on CPUs to the best of our knowledge. Our technique is orthogonal to and may be combined with other Transformer optimization techniques such as GEMM optimizations (e.g., \cite{dice2021optimizing}).

\paragraph{Intel AMX Extension}
Intel's AMX \cite{intel2023amx} is designed to accelerate matrix-level operations on CPUs, and only available in high-performance processors like the 4th Generation Xeon Scalable Processors \cite{intel2023xeonscalable}. Our research focuses on prevalent SIMD extensions. Moreover, it is essential to develop dataflows that maximize data reuse opportunities in AMX to further optimize its performance, and our methodology may be extended for this purpose.

\paragraph{Binary Neural Network Optimizations}
Frameworks that optimize binary neural networks specifically exist. An example is daBNN~\cite{zhang2019dabnn}, which employs various assembly-level microkernels to optimize performance. However, daBNN fails to harvest all data reuse opportunities, such as reusing input data between two successive outputs, or reusing weight data. By combining our dataflow technique with daBNN, further improvements can be achieved.     

\section{Conclusions}
In this paper, we present the first study to systematically explore dataflows to achieve efficient neural network inference using SIMD capabilities. We 
developed heuristics for optimized vector register allocation by analyzing reuse opportunities for different dataflows, and validated these heuristics through thorough automatic code generation and experimentation, and demonstrated significant performance improvements over state-of-the-art implementations. 
We anticipate that this work will catalyze further investigation of dataflows to reduce inference time on contemporary CPU architectures.


\bibliography{reference}





\end{document}